\documentclass[twocolumn,showpacs,preprintnumbers,amsmath,amssymb]{revtex4}

\usepackage{graphicx}% Include figure files
\usepackage{dcolumn}% Align table columns on decimal point
\usepackage{bm}% bold math
\usepackage{booktabs}

\begin{document}
\preprint{APS}
\title{Miniaturized saturated absorption spectrometer}
\author{K. Sosa}
\author{J. Oreggioni}
\affiliation{Instituto de Ingenier\'ia El\'ectrica, Facultad de Ingenier\'ia, Universidad de la Rep\'ublica,\\ J. Herrera y Reissig 565, 11300 Montevideo, Uruguay}
\author{H. Failache} \email{heraclio@fing.edu.uy}
\affiliation{Instituto de F\'isica, Facultad de Ingenier\'ia, Universidad de la Rep\'ublica,\\ J. Herrera y Reissig 565, 11300 Montevideo, Uruguay}
\date{\today}

\begin{abstract}
We describe a saturated absorption setup that is robust, compact and require minimum alignment. These properties are attained using a diffuse probe beam generated by a retro-reflecting film. This concept was studied and applied to built a miniaturized setup that had shown the same performance than an optimized table-top experiment.
\end{abstract}


%42.62.Fi Laser spectroscopy

\maketitle
\section{\label{Introduction}Introduction}
Laser sources integrated in miniaturized devices like atomic clocks, magnetometers and gyroscopes are typically locked to an atomic transition to achieve a good frequency stability \cite{Kitching:18}. Most of the approaches to stabilize a laser source to an atomic transition are based on the two-photon Doppler-free spectroscopy method initially proposed by Vasilenko et al. \cite{Vasilenko:70}, usually referred as saturated absorption (SA) spectroscopy \cite{Demtroder:96}. This method use two counter-propagating laser beams in a thermal atomic vapor. In this configuration, the atoms that are simultaneously resonant to both beams are those that are not Doppler shifted, i.e. those with velocities perpendicular to the beams. Using this velocity-selective non-linear interaction, narrow Doppler-free resonances are observed at the resonance frequency of atoms at rest. A residual Doppler broadening may be nevertheless introduced if a misalignment out of the counter-propagating beams configuration is not controlled.\\
\begin{figure}
\includegraphics[width=9cm]{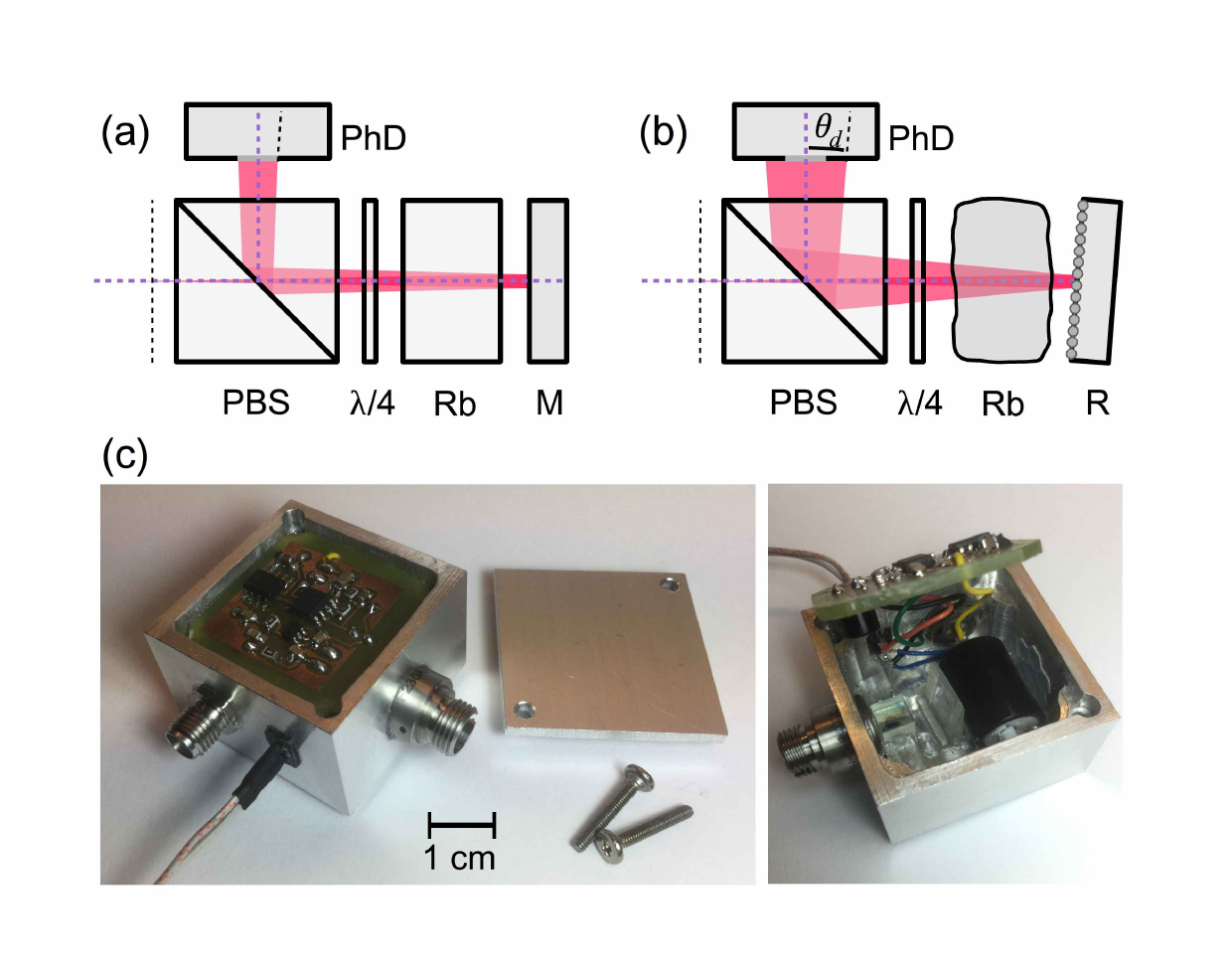}
\caption{(Color online) (a) Simple scheme of saturated absorption spectroscopy, (b) proposed setup using a diffuse light field probing the atomic medium. (PBS: polarizing-beam-splitter, $\lambda$/4: quarter wave plate, Rb: rubidium cell, M: mirror, R: reflector, PhD: photodetector), (c) miniaturized setup.} \label{setup}
\end{figure}
Some compact \cite{Affolderbach05} or miniaturized setups \cite{Gerginov:06,Knappe:07,Groswasser09} have been proposed to stabilize a laser source frequency using SA resonances. Inspired by the work of Villalba et al. \cite{Villalba:14}, who observed SA resonances on an atomic vapor probed by diffuse light, we consider in this work the possibility of using diffuse light in a SA setup and we identify some advantages for its miniaturization.\\
The SA configuration shown in Fig.1(a) has the virtue of using a small number of components, without compromising the basic requirement of having the pump and probe beams interacting with the atomic vapor in perfect contra-propagating directions. Additionally, if the quarter wave-plate position is chosen to define circularly polarized light in the vapor cell, after reflection in the mirror and a second passage through the wave-plate, its polarization becomes linear and orthogonal to the incident polarization. The light with such polarization is then reflected in the polarizing beam-splitter (PBS) reducing the transmitted probe light that is usually wasted when a standard beam splitter is used instead.\\
We have considered however the setup shown in Fig.1(b), where the probe beam is replaced by diffuse laser radiation that is obtained substituting the mirror by a light scattering surface. We show that even if a small broadening is introduced, the setup becomes robust, insensitive to any misalignment or movement of the diffusing surface that does not require any alignment. Moreover, we propose to use an engineered diffusing surface, like a retro-reflective film, that brings additional advantages, like a higher probe light intensity and the possibility to correct a misalignment introduced by vapor cell windows which could eventually have a poor optical quality.\\

\begin{figure}
\includegraphics[width=8cm]{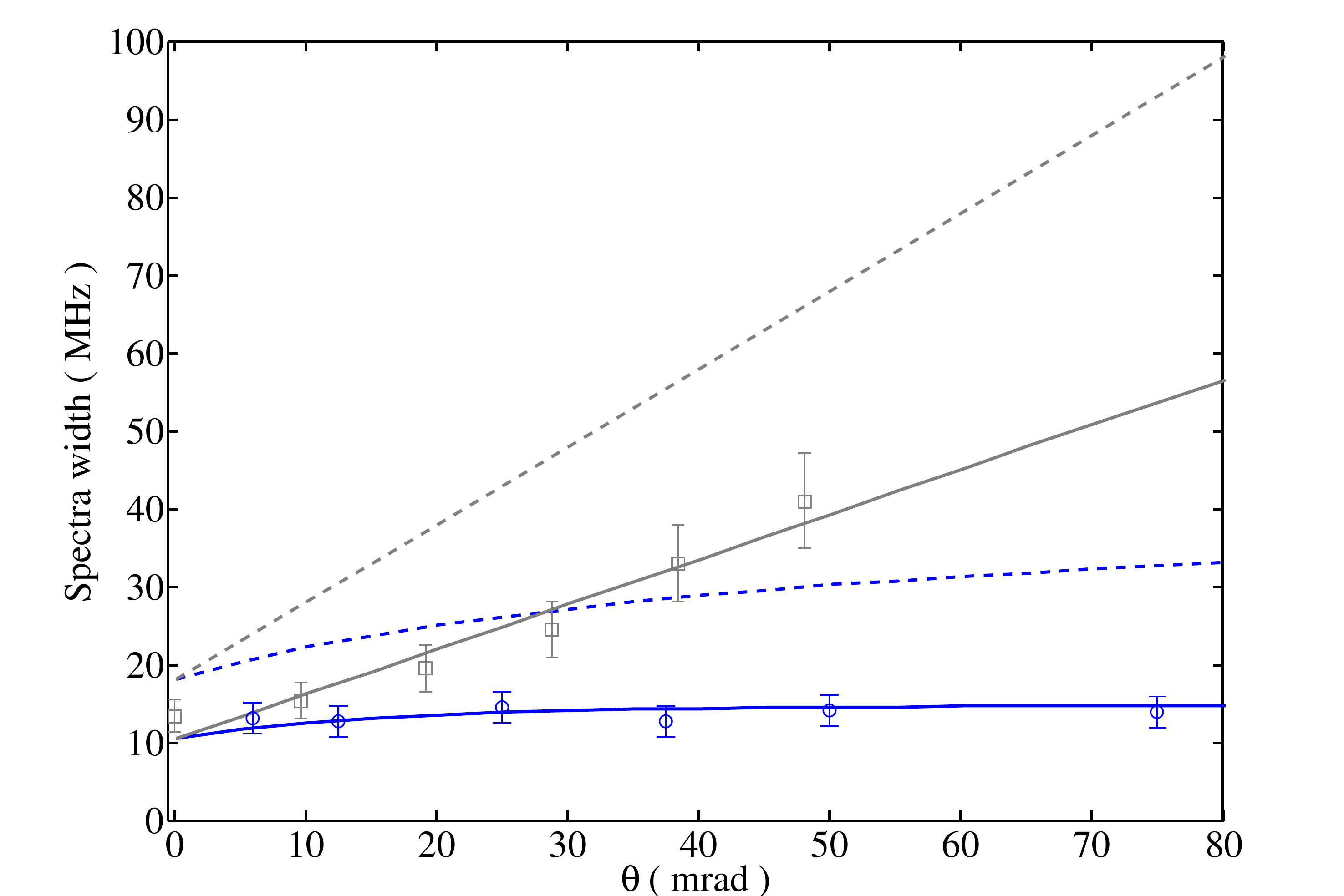}
\caption{(Color online) FWHM of the SA spectra for a diffuse probe radiation with a characteristic spreading angle $\theta_d$ (dashed blue) and peak-to-peak width $\Delta_{pp}$ of the derivative of same spectra (continuous blue), after the simplified theoretical model described in the main text. The dependence of the SA for probe and pump beams with an angle 2$\theta_{pp}$ between them (dashed grey) and their derivative (continuous grey) are also shown for comparison. Experimental measurements for both cases are shown. These measurements are affected by a small power broadening considered in the model.} \label{broaden}
\end{figure} 

\section{Diffuse probe beam}

In the standard SA setup of Fig.1(a), a careful alignment of the mirror is required to have perfectly counter-propagating pump and probe beams in order to avoid a residual Doppler broadening. We propose to replace this mirror by a scattering surface, which scatters light over a large solid angle around the back-reflected direction. Therefore this scattering surface generates a probe light along the much narrower detection solid angle almost independently of its orientation, turning its careful alignment unnecessary. However, this advantage is counterbalanced by two main consequences of using diffuse probe light. First, the angular dispersion of probe light may introduce a residual Doppler broadening and second, the amount of light in the small detection solid angle may be low, reducing the signal amplitude and its signal-to-noise ratio. As we consider below, these two side effects can be strongly reduced.\\

To quantify the Doppler broadening, we describe the probe light as a field that is confined within a dispersion cone with a half-apex angle $\theta_d$. We consider a probe radiation with a wave-vector dispersion on the range (-$\theta_d$,$\theta_d$) and we model the SA spectra as the cumulative contribution of spectra corresponding to pump and probe wave-vectors having angles in this range. Using this simplified description, the SA spectrum is therefore calculated as an integral, over the pump-probe angle, of Lorentzian curves each having a width affected by the corresponding residual Doppler broadening and constant amplitude. In Fig.2 we show the full-width at half-maximum (FWHM) of the resulting SA spectrum, and the peak-to-peak width $\Delta_{pp}$ for their derivative, as a function of $\theta$=$\theta_d$.  We have considered the derivative because this is the useful signal in most servo-control system, where the resulting dispersive spectrum is used as the error signal to control the laser source wavelength. In Fig.2 we also show, for comparison, SA spectra width for a probe beam having an angle $\theta$=2$\theta_d$ with the pump. Experimental measurements are also shown for the corresponding derivated spectra. The diffuse probe light for these measurements was generated by the scattering surface of a white paper.\\ 
\begin{figure}
\includegraphics[width=9cm]{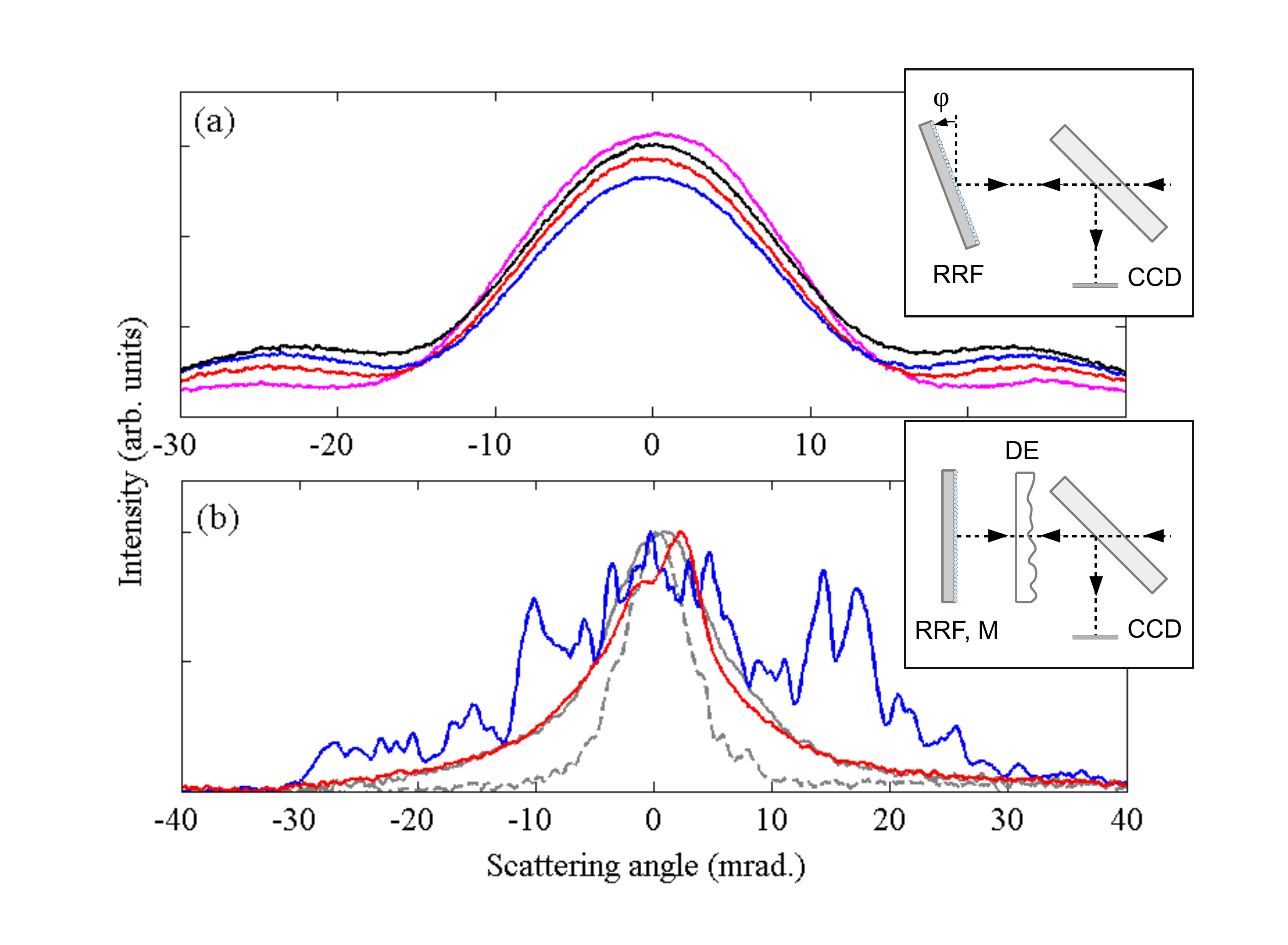}
\caption{(Color online) (a) Light distribution profile of the scattered radiation of the RRF for different incidence angles $\varphi$ of the light beam (magenta: $\varphi$ = 8$^o$, red: $\varphi$ = 14$^o$, blue: $\varphi$ = 24$^o$ , black: $\varphi$ = 34$^o$). (b) Light distribution profile at the position of the detector in Fig.1(b) when a distorting element (DE), like an uneven glass surface, is placed before a plane mirror M (blue) or a RRF (red). The undistorted light distribution profile corresponding to a plane mirror (dashed grey) and a RRF (continuos grey) are also shown for comparison. All curves were normalized.} \label{rrf}
\end{figure} 
Notably, using a diffuse probe field, only a small residual Doppler broadening of few MHz is observed on the $\Delta_{pp}$ width of the derivative of the SA spectra for any angle $\theta_d$. This fact can be simply explained considering that in a spectrum resulting of a cumulative contribution of spectra of different widths, narrower spectra have a much larger contribution to the derivative than broader ones.\\

Considering now the wasted probe light that is scattered out of the detection solid angle, it can be substantially reduced if the scattering surface in Fig.1(b) is replaced by a retro-reflective film (RRF) \cite{Hericz:17}. These films, typically employed in traffic or safety signaling, scatters light within a narrow cone around the incidence direction, which strongly reduce the probe power lost outside of the photodetector. Moreover, the RRFs back-scatters light almost independently of the incidence angle, as can be seen in Fig.3(a), where we show the intensity distribution profile of the light at the detector position after reflection on a RRF for different incidence angles.\\
The fact that the light is retro-reflected by a RRF is a fundamental difference with respect to a diffusive surface. An important consequence is that a ray deviation generated by a distorting medium in the path of the light incident on a RRF, is essentially corrected after the reflection on this film and a second passage through the medium in the opposite direction. In Fig.3(b) we show the radiation distribution profile at the detector position, when an uneven glass surface is placed in front of the reflective element. The distorted profile obtained when the reflective element is a mirror (blue curve in Fig.3(b)) is essentially corrected when a RRF is used instead (red curve in Fig.3(b)). Using a RRF in the setup shown in Fig.1(b) makes it therefore quite insensitive to any movement or misalignment generated by the atomic vapor cell or the RRF itself.\\
It is however important to consider that RRFs are constructed with micro-reflecting elements that typically have a size of several tens of micrometers. A light beam incident on these RRF is reflected on a direction that is almost parallel to the incidence one but slightly shifted by a distance that is of the order of the reflecting element size. These RRF are therefore only able to make this correction on beam-front distortions at scales larger than the micro-reflector element size, and under the assumption that light propagation can be described using only geometric optics.\\

\section{\label{Introduction}Miniaturized setup}

In Fig.1(c) we show photographs of a miniaturized setup, corresponding to the scheme shown in Fig.1.(b), encapsulated in a metallic box measuring 3.2 cm x 3.2 cm x 2.4 cm. This setup can be further reduced in size if smaller optical elements are employed. The laser light arrives to the setup through a polarization-maintaining single mode optical fiber and is collimated with a lens to a light beam of around 1 mm in diameter. We have used a 5 mm PBS cube and a small glass cell filled with pure Rb vapor fabricated using a novel miniaturization technique that will be reported elsewhere. The cell has an approximate external volume of 100 mm$^3$ with a light path in the Rb vapor of around 1 mm. To increase the optical density of the atomic medium, the Rb density was increased heating the cell to 80 Celsius using an electrical wire wrapped around it. Larger densities are however possible increasing the power consumption of the device. We have used a 3M Scotchlite reflective material, which retro-reflect 21 $\%$ of incident light power within a narrow cone. Their retro-reflected radiation profile is shown in Fig.3(a), where it can be seen that the divergence angle $\theta_d$ is around 10 mrad. We have used a photodiode with a square sensing area of 1 mm on a side, placed at a distance of approximately 20 mm to the RRF. The half-apex angle of the photodetector collecting cone is then approximately twice the divergence angle $\theta_d$, which means that a small fraction of the probe power is wasted.\\
\begin{figure}
\includegraphics[width=8cm]{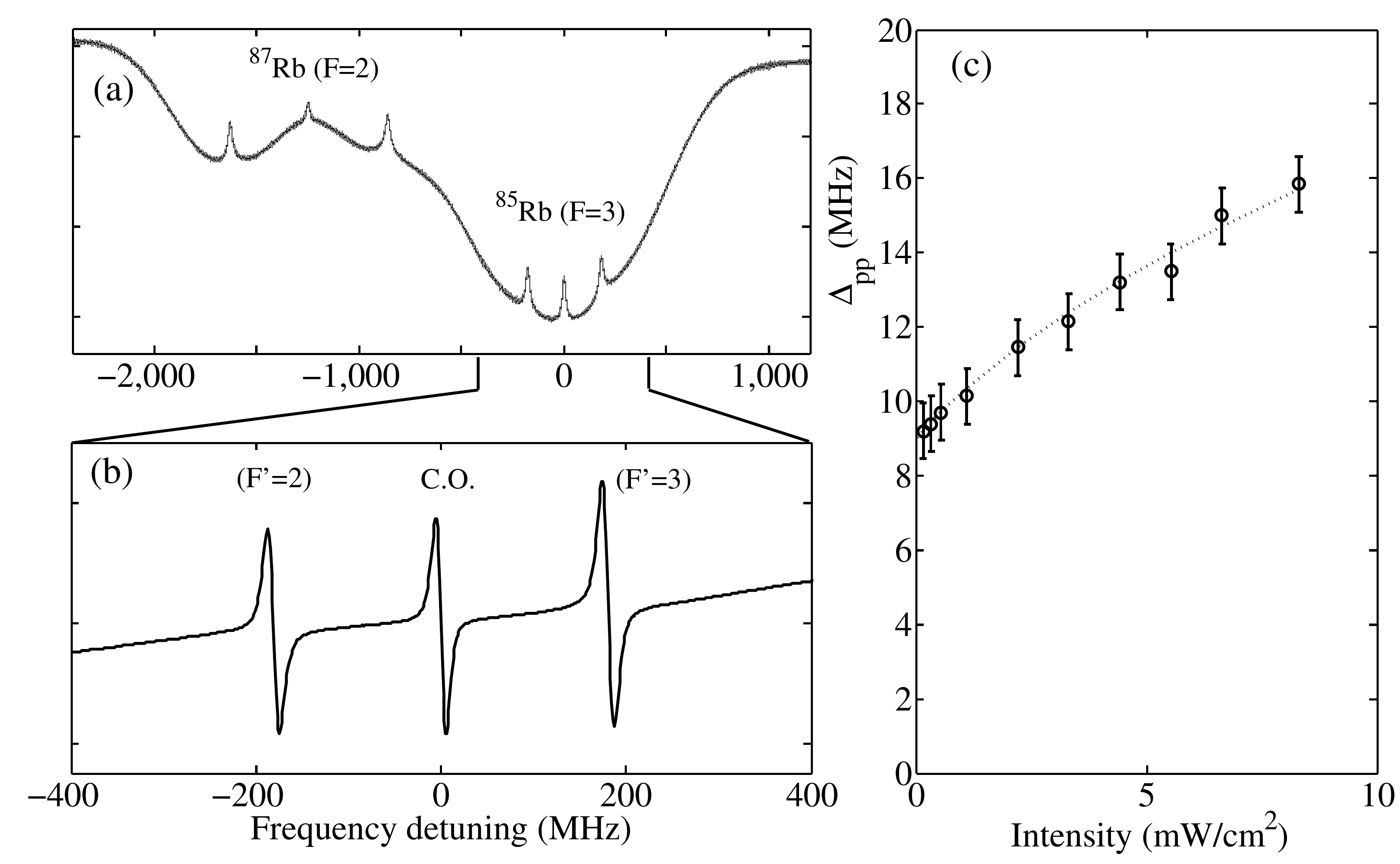}
\caption{(a) Example of a SA spectrum measured directly on the scope for the miniaturized setup (using 200 $\mu$W input light power, averaged over 64 spectra), (b) synchronically detected SA spectrum, (c) Peak-to-peak SA spectra width $\Delta_{pp}$ as a function of the input light intensity (dashed line: simple model fitted to the experimental data \cite{Akulshin:1990}).} \label{broaden}
\end{figure}
Concerning the optical alignment of the setup in Fig.1(b), a given direction of the incoming pump beam uniquely defines the path of the contra-propagating beams. The PBS position, which can be independently defined, determines the direction along which the probe is reflected and hence the photo-detector position. The mechanics were then designed to hold in place the fiber collimator, the PBS and the photo-diode with such a precision that any alignment of the optical system becomes unnecessary. According to the discussion above, the detection solid angle does not determine the width of SA spectra and is not a relevant parameter. Moreover, the precision required to place the detector along the counter-propagating direction is largely reduced using a photodiode with a relatively large detection area. A mechanical uncertainty of 100 $\mu$m, easily achieved in any mechanical piece, is large enough to have a good optical alignment of our setup.\\

The SA signal is detected and pre-amplified inside the metallic box with a low-cost amplifier, made with off-the-shelf components, carefully designed to comply with low noise requirements. We have used the OSRAM SFH 203 PFA pin photodiode (dark current $\sim$ 1 nA, responsivity $\sim$ 0.6 A/W at $\lambda$ = 795 nm). The circuitry is powered by an external supply of 8V, that also powers the cell heater. The total power consumption is 600 mW. A coaxial electrical plug connects the SA output signal directly to external electronics. The electronics of photo-detection have a Noise Equivalent Power = 0.3 pW/$\sqrt Hz$.\\
In Fig.4(a) we show an example of a SA absorption spectra measured at the output of our miniaturized setup directly with an oscilloscope. We have used an extended cavity diode laser resonant with the $^{85}$Rb: 5S$_{1/2}(F=3) \rightarrow$ 5 P$_{1/2}$(F'=2,3) and $^{87}$Rb: 5S$_{1/2}(F=2) \rightarrow$ 5 P$_{1/2}$(F'=1,2) transitions. We have also used a frequency modulated laser and a locking detection to measure the frequency derivated spectrum shown in Fig.4(b). For an input light power of 50 $\mu$W and 100 ms integrating time constant we have measured a signal-to-noise ratio (S/N) of approximately 400. In Fig.4(c) we show the peak-to-peak width $\Delta_{pp}$ of these frequency derivated SA spectra as a function of the input laser intensity. The width $\Delta_{pp}$ limit for low intensity is 9.0 $\pm$ 0.5 MHz. The same minimum width value was measured in a table-top SA setup, where we have taken special care to reduce the effect of broadening mechanisms like power broadening, pump-probe beams misalignment and atomic collisions. It should be mentioned that we were unable to measure a $\Delta_{pp}$ closer to the natural width of the probed transition (5.8 MHz) due to some frequency noise in the diode laser.\\
Considering the SA spectra as an error signal used to stabilice the frequency of a laser source, the relevant parameter to take into account is the slope around the control frequency. This parameter was particularly not optimized in the spectra shown previously. We have verified that as the light intensity is increased this slope monotonically grows, as well as the S/N of the signal, up to the maximum intensity considered in this work.\\

In conclusion, we have studied a SA setup configuration that is robust, compact and alignment-free. A miniaturized setup was built, as a proof-of-principle, that shows the same performance than an optimized SA table-top setup.\\

\section{Acknowledgments}

We gratefully acknowledge fruitful discussions with Arturo Lezama and financial support from CSIC (Universidad de la Rep\'ublica).\\

%\bibliography{ReferenciasSA19}

\begin{thebibliography}{10}
\expandafter\ifx\csname natexlab\endcsname\relax\def\natexlab#1{#1}\fi
\expandafter\ifx\csname bibnamefont\endcsname\relax
  \def\bibnamefont#1{#1}\fi
\expandafter\ifx\csname bibfnamefont\endcsname\relax
  \def\bibfnamefont#1{#1}\fi
\expandafter\ifx\csname citenamefont\endcsname\relax
  \def\citenamefont#1{#1}\fi
\expandafter\ifx\csname url\endcsname\relax
  \def\url#1{\texttt{#1}}\fi
\expandafter\ifx\csname urlprefix\endcsname\relax\def\urlprefix{URL }\fi
\providecommand{\bibinfo}[2]{#2}
\providecommand{\eprint}[2][]{\url{#2}}

\bibitem[{\citenamefont{Kitching}(2018)}]{Kitching:18}
\bibinfo{author}{\bibfnamefont{J.}~\bibnamefont{Kitching}},
  \bibinfo{journal}{Applied Physics Reviews} \textbf{\bibinfo{volume}{5}},
  \bibinfo{pages}{031302} (\bibinfo{year}{2018}).

\bibitem[{\citenamefont{Vasilenko et~al.}(1970)\citenamefont{Vasilenko,
  Chebotayeb, and Shishaev}}]{Vasilenko:70}
\bibinfo{author}{\bibfnamefont{L.~S.} \bibnamefont{Vasilenko}},
  \bibinfo{author}{\bibfnamefont{V.~P.} \bibnamefont{Chebotayeb}},
  \bibnamefont{and} \bibinfo{author}{\bibfnamefont{A.~V.}
  \bibnamefont{Shishaev}}, \bibinfo{journal}{JETP Lett.}
  \textbf{\bibinfo{volume}{12}}, \bibinfo{pages}{113} (\bibinfo{year}{1970}).

\bibitem[{\citenamefont{Demtroder}(1996)}]{Demtroder:96}
\bibinfo{author}{\bibfnamefont{W.}~\bibnamefont{Demtroder}},
  \emph{\bibinfo{title}{Laser Spectroscopy}}
  (\bibinfo{publisher}{Springer-Verlag}, \bibinfo{year}{1996}).

\bibitem[{\citenamefont{Affolderbach and Mileti}(2005)}]{Affolderbach05}
\bibinfo{author}{\bibfnamefont{C.}~\bibnamefont{Affolderbach}}
  \bibnamefont{and} \bibinfo{author}{\bibfnamefont{G.}~\bibnamefont{Mileti}},
  \bibinfo{journal}{Review of Scientific Instruments}
  \textbf{\bibinfo{volume}{76}}, \bibinfo{pages}{073108}
  (\bibinfo{year}{2005}).

\bibitem[{\citenamefont{Gerginov et~al.}(2006)\citenamefont{Gerginov, Shah,
  Knappe, Hollberg, and Kitching}}]{Gerginov:06}
\bibinfo{author}{\bibfnamefont{V.}~\bibnamefont{Gerginov}},
  \bibinfo{author}{\bibfnamefont{V.}~\bibnamefont{Shah}},
  \bibinfo{author}{\bibfnamefont{S.}~\bibnamefont{Knappe}},
  \bibinfo{author}{\bibfnamefont{L.}~\bibnamefont{Hollberg}}, \bibnamefont{and}
  \bibinfo{author}{\bibfnamefont{J.}~\bibnamefont{Kitching}},
  \bibinfo{journal}{Opt. Lett.} \textbf{\bibinfo{volume}{31}},
  \bibinfo{pages}{1851} (\bibinfo{year}{2006}).

\bibitem[{\citenamefont{Knappe et~al.}(2007)\citenamefont{Knappe, Robinson, and
  Hollberg}}]{Knappe:07}
\bibinfo{author}{\bibfnamefont{S.~A.} \bibnamefont{Knappe}},
  \bibinfo{author}{\bibfnamefont{H.~G.} \bibnamefont{Robinson}},
  \bibnamefont{and} \bibinfo{author}{\bibfnamefont{L.}~\bibnamefont{Hollberg}},
  \bibinfo{journal}{Opt. Express} \textbf{\bibinfo{volume}{15}},
  \bibinfo{pages}{6293} (\bibinfo{year}{2007}).

\bibitem[{\citenamefont{Groswasser et~al.}(2009)\citenamefont{Groswasser,
  Waxman, Givon, Aviv, Japha, Keil, and Folman}}]{Groswasser09}
\bibinfo{author}{\bibfnamefont{D.}~\bibnamefont{Groswasser}},
  \bibinfo{author}{\bibfnamefont{A.}~\bibnamefont{Waxman}},
  \bibinfo{author}{\bibfnamefont{M.}~\bibnamefont{Givon}},
  \bibinfo{author}{\bibfnamefont{G.}~\bibnamefont{Aviv}},
  \bibinfo{author}{\bibfnamefont{Y.}~\bibnamefont{Japha}},
  \bibinfo{author}{\bibfnamefont{M.}~\bibnamefont{Keil}}, \bibnamefont{and}
  \bibinfo{author}{\bibfnamefont{R.}~\bibnamefont{Folman}},
  \bibinfo{journal}{Review of Scientific Instruments}
  \textbf{\bibinfo{volume}{80}}, \bibinfo{pages}{093103}
  (\bibinfo{year}{2009}).

\bibitem[{\citenamefont{Villalba et~al.}(2014)\citenamefont{Villalba, Laliotis,
  Lenci, Bloch, Lezama, and Failache}}]{Villalba:14}
\bibinfo{author}{\bibfnamefont{S.}~\bibnamefont{Villalba}},
  \bibinfo{author}{\bibfnamefont{A.}~\bibnamefont{Laliotis}},
  \bibinfo{author}{\bibfnamefont{L.}~\bibnamefont{Lenci}},
  \bibinfo{author}{\bibfnamefont{D.}~\bibnamefont{Bloch}},
  \bibinfo{author}{\bibfnamefont{A.}~\bibnamefont{Lezama}}, \bibnamefont{and}
  \bibinfo{author}{\bibfnamefont{H.}~\bibnamefont{Failache}},
  \bibinfo{journal}{Phys. Rev. A} \textbf{\bibinfo{volume}{89}},
  \bibinfo{pages}{023422} (\bibinfo{year}{2014}).

\bibitem[{\citenamefont{H\'{e}ricz et~al.}(2017)\citenamefont{H\'{e}ricz,
  Sarkadi, Erdei, Lazuech, Lenk, and Koppa}}]{Hericz:17}
\bibinfo{author}{\bibfnamefont{D.}~\bibnamefont{H\'{e}ricz}},
  \bibinfo{author}{\bibfnamefont{T.}~\bibnamefont{Sarkadi}},
  \bibinfo{author}{\bibfnamefont{G.}~\bibnamefont{Erdei}},
  \bibinfo{author}{\bibfnamefont{T.}~\bibnamefont{Lazuech}},
  \bibinfo{author}{\bibfnamefont{S.}~\bibnamefont{Lenk}}, \bibnamefont{and}
  \bibinfo{author}{\bibfnamefont{P.}~\bibnamefont{Koppa}},
  \bibinfo{journal}{Appl. Opt.} \textbf{\bibinfo{volume}{56}},
  \bibinfo{pages}{3969} (\bibinfo{year}{2017}).

\bibitem[{\citenamefont{Akulshin et~al.}(1990)\citenamefont{Akulshin,
  Sautenkov, Velichansky, Zibrov, and Zverkov}}]{Akulshin:1990}
\bibinfo{author}{\bibfnamefont{A.}~\bibnamefont{Akulshin}},
  \bibinfo{author}{\bibfnamefont{V.}~\bibnamefont{Sautenkov}},
  \bibinfo{author}{\bibfnamefont{V.}~\bibnamefont{Velichansky}},
  \bibinfo{author}{\bibfnamefont{A.}~\bibnamefont{Zibrov}}, \bibnamefont{and}
  \bibinfo{author}{\bibfnamefont{M.}~\bibnamefont{Zverkov}},
  \bibinfo{journal}{Optics Communications} \textbf{\bibinfo{volume}{77}},
  \bibinfo{pages}{295} (\bibinfo{year}{1990}).

\end{thebibliography}

\end{document}